\documentstyle[12pt,epsf]{article}

\begin{document}

\title{Orbital effects in manganites\footnote{This paper is dedicated
to Prof.~E.~M\"uller-Hartmann on the occasion of his 60th birthday}}

\author{D.~I.~Khomskii\\ \\
Laboratory of Solid State Physics,\\Groningen University,\\
Ni\kern-0.07em jenborgh~4, 9747~AG~Groningen,\\
The~Netherlands}

\date{ }

\maketitle

\begin{abstract}
In this paper I give a short review of some properties of the colossal 
magnetoresistance manganites, connected with the orbital degrees of 
freedom. Ions Mn$^{3+}$, present in most of these compounds, have 
double orbital degeneracy and are strong Jahn-Teller ions, causing 
structural distortions and orbital ordering. Mechanisms leading to 
such ordering are shortly discussed, and the role of orbital degrees 
of freedom in different parts of the phase diagram of manganites is 
described. Special attention is paid to the properties of low-doped 
systems (doping $0.1\leq x\leq0.25$), to overdoped systems ($x > 0.5$), 
and to the possibility of a novel type of orbital ordering in optimally 
doped ferromagnetic metallic manganites.
\end{abstract}

\section{Introduction}
When considering the properties of real systems with strongly
correlated electrons, such as transition metal (TM) oxides, one often
has to take into account, besides the charge and spin degrees of
freedom, described e.g.\ by the nondegenerate Hubbard model, also the
orbital structure of corresponding TM ions.  These orbital degrees of
freedom are especially important in cases of the so-called orbital
degeneracy---the situation when the orbital state of the TM ions in
a regular, undistorted coordination (e.g.\ in a regular
$O_6$-octahedron) turns out to be
degenerate~\cite{goodenough,kugel,nagaosa}. This is e.g.\ the
situation with the ions Cu$^{2+}$ ($d^9$), Mn$^{3+}$ ($d^4$),
Cr$^{2+}$ ($d^4$), low-spin Ni$^{3+}$ ($d^7=t_{2g}^6e_g^1$). This
degeneracy gives rise in an isolated centre to the famous Jahn-Teller
effect~\cite{jahn}, and in concentrated systems---to the
cooperative transition which may be viewed as simultaneous structural
phase transition lifting this orbital degeneracy (cooperative
Jahn-Teller transition), as an orbital ordering (OO) and as a
quadrupolar ordering (this latter terminology is often used in rare
earth compounds).

All these effects play very important role in the materials which
became very popular recently---in manganites with the colossal
magnetoresistance (CMR).  The typical example is the system
La$_{1-x}$Ca$_x$MnO$_3$ (there may be other rare earths, e.g.\ Pr, Nd,
or Bi instead of La, or other divalent cations---Sr, Ba, Pb---instead
of Ca; there exist also layered materials of this kind).  The typical
phase diagram of these systems is shown schematically on
fig.~1.  The starting undoped material LaMnO$_3$, which is an
antiferromagnetic insulator, contains  typical Jahn-Teller ions
Mn$^{3+}$ (electronic configuration $t_{2g}^3e_g^1$)---i.e.\ it is
orbitally doubly-degenerate, see fig.~2.   Thus we can expect
that the orbital degrees of freedom may significantly influence the
properties of CMR manganites---the idea which is largely supported
by experiments.  In this short paper I will try to review some
of the aspects of the physics of manganites connected with orbital
degrees of freedom. This is actually
already quite large and well developed
field,  and  of  course  I will not be
able to cover this whole field;  much
of  the  material  presented  will  be
based on the investigations in which I
myself  participated.   Some  of   the
general concepts  used below  are also
presented in~\cite{khomskii,khomskii2}.

This paper is dedicated to
Prof.~Erwin M\"uller-Hartmann on the occasion of
his 60-th birthday.  During many years of my
acquaintance with him, especially
during my stay in Cologne
(1990--1992) I benefited a lot from many
contacts with him.  Also lately we had
very interesting and fruitful
discussions in connection with my recent, still controversial idea
about ``complex orbitals'' ordering which is also presented below in
this paper.  I am happy to use this opportunity to wish him all the
best in the years to come.

\section{Main features of the phase diagram of manganites: orbital
effects}
As already mentioned above, the undoped LaMnO$_3$ with the perovskite
structure contains strong Jahn-Teller ions Mn$^{3+}$.  They are known
always to induce rather strong local distortion in all the insulating
compounds containing them~\cite{kugel}.  Also here it is well
known that there exists an orbital ordering and concomitant lattice
distortion in LaMnO$_3$: $e_g$-orbitals of Mn$^{3+}$ ions are
ordered in such a way that at the neighbouring Mn sites the
alternating $x^2$ and $y^2$-orbitals are occupied, i.e.\ the local
O$_6$-octahedra are alternatingly elongated along $x$ and
$y$-directions, see fig.~3.

Orbital ordering is also known to exist in most manganites in another
well-defined region of the phase diagram---at half-doping $x=0.5$.  In
this situation there occurs with decreasing temperature the charge
ordering---the checkerboard arrangement of Mn$^{3+}$ and Mn$^{4+}$
ions in the basal plane, see
fig.~4~\cite{wollan,jirak,goodenough}.   The Mn$^{3+}$-ions
with localized electrons again have an orbital degeneracy (Mn$^{4+}$
ions are nondegenerate, cf.~fig.~2) and develop the orbital
ordering shown in fig.~4.  Both the charge ordering (CO) and
the OO occur simultaneously at the same temperature, although from
some data it follows that probably the CO is the driving force, and
the OO follows it~\cite{zimmermann}; this however is still an open
question.

According to the well-known Goodenough--Kanamori--Anderson rules (see
e.g.~\cite{goodenough,khomskii,khomskii2}) the magnitude and even the
sign of the magnetic exchange depend on the type of orbitals
occupied.  Thus if the occupied by one electron (half-filled) orbitals
are directed towards one another, one has a strong antiferromagnetic
coupling; if however these orbitals are directed away from each other
(are mutually orthogonal, as e.g.\ in fig.~3) we would
have a ferromagnetic interaction.  That is why the undoped LaMnO$_3$
(fig.~3) has the A-type magnetic ordering---the spins in the
$(x,y)$-plane order ferromagnetically, the next $xy$-layer being
antiparallel to the first one.

Similarly, the exchange between Mn$^{3+}$ and Mn$^{4+}$ ions is
ferromagnetic if the orbital of Mn$^{3+}$ ion is directed towards
Mn$^{4+}$ and antiferromagnetic if it looks away from it.  This gives
rise to the very specific CE-type magnetic structure for $x=0.5$
manganites, shown in fig.~4: we have here the ferromagnetic
zigzags, stacked antiferromagnetically. Recall now that according to
the double-exchange picture (\cite{degennes}, see
also~\cite{khomskii})  electrons can easily move on the
ferromagnetic background, but  electron hopping is forbidden or at
least strongly suppressed if the spins are antiparallel.  Then one
immediately sees that the motion of electrons in the
CE-background shown in fig.~4 becomes essentially
one-dimensional---it is confined to the ferromagnetic zigzags shown
in this figure~\cite{soloviev,vandenbrink}. One may  show that due to
a special topology of these zigzags (the presence of the edge and
corner sites, different orbitals occupation), the gap opens in the
1$d$ tight-binding spectrum of electrons even if we do not assume the
charge ordering~\cite{soloviev,vandenbrink}.  Charge ordering
appears when we add in this situation not even the intersite Coulomb
repulsion (which would definitely prefer to stabilize the CO state),
but the on-site Hubbard repulsion~\cite{vandenbrink}.  What is the
actual mechanism leading to the CE-type (charge, orbital and
magnetic) ordering in $x=0.5$ manganites, will be discussed in the
next section.

There are two more regions of the phase diagram of fig.~1 in
which  orbital effects apparently play an important role, although
the detailed picture is less clear.  These are the low-doped region
$0.1\leq x\leq0.2$ -- $0.3$ (depending on the specific system
considered) in which one often observes ferromagnetic insulating
(FI) and often charge-ordered phase.  This is the case of
La$_{1-x}$Ca$_x$MnO$_3$ ($0.1\leq x\leq0.25$),
La$_{1-x}$Sr$_x$MnO$_3$ close to $x=\frac18$ ($0.1\leq x\leq0.18$)~\cite{klingerer,endo} or
Pr$_{1-x}$Ca$_x$MnO$_3$ ($0.15\leq x\leq0.3$)~\cite{jirak}.

It is rather uncommon to have the FI state: typically insulating
material of this class are antiferromagnetic, and ferromagnetism goes
hand in hand with metallicity which finds natural explanation in
the model of double exchange~\cite{degennes}.  The only possibility
to obtain the FI state in perovkites is due to a certain particular
orbital ordering favourable for ferromagnetism~\cite{khomskii} (FI
state can appear also in systems in which there exists the
$90^\circ$-superexchange---the TM--O--TM angle is close to
$90^\circ$).  But what is the detailed ordering in this low-doped
region, is not completely clear, see below.

Another interesting, and much less explored, region is the overdoped
manganites, $x>0.5$.  Typically in this case we have an insulating
state, sometimes with the CO and OO state in the form of
stripes~\cite{radaelli} or bistripes~\cite{mori}, see
fig.~5.

The choice between these two options is still a  matter of
controversy (see e.g.~\cite{radaelli,wang}), as well as the detailed type of magnetic
ordering in this case.  I will return to this point in sec.~5; in any
case we see that in this stripe-like phase the orbital degrees of
freedom definitely play an important, and maybe decisive role---see
sec.~5.

Returning to magnetic properties of overdoped manganites $x>0.5$,
one should mention an important fact: the very strong asymmetry of the
typical phase diagram of manganites.  As seen e.g.\ from
fig.~1, there usually exists a rather large ferromagnetic
metallic region (FM) for $x<0.5$, but nearly never for $x>0.5$ (only
rarely does one observe bad metal and unsaturated ferromagnetism in a
narrow concentration range in some overdoped
manganites~\cite{martin}).  However from the standard
double-exchange model one can expect the appearance of a FM phase not
only in hole-doped LaMnO$_3$ ($x<0.5$) but in electron-doped
CaMnO$_3$ ($x>0.5$) as well. Orbital degeneracy may play some
role in explaining this asymmetry~\cite{vandenbrink2}---see sec.~5.

There exists also a problem what are the orbitals doing in the optimally
doped ferromagnetic and metallic manganites. Usually one completely
ignores orbital degrees of freedom in this regime, at least at low
temperatures; this is supported by the experimental observations
that the MnO$_6$-octahedra are completely regular in this case. There
is however a possibility that there still exists in this case an
orbital ordering, but of completely novel type, not accompanied by
any lattice distortion---an ordering of complex orbitals~\cite{khomskii4,vandenbrink3}.
This, still rather controversial point will be discussed in sec.~6.

Finally, one can also
ask a question---which role do orbital degrees of freedom and
Jahn-Teller play at elevated temperature, in particular in
disordered states. These questions will be shortly discussed in
sec.~7.

\section{Mechanisms of Orbital Ordering}
Before discussing particular situations in different doping
ranges, it is worthwhile to shortly discuss the general question of
possible interactions of degenerate orbitals which can
lead to orbital ordering.  In transition metal compounds there are
essentially two such mechanisms.  The first one is connected with the
Jahn-Teller interaction of degenerate orbitals with the lattice
distortions, see e.g.~\cite{gehring}.  Another mechanism was proposed
in 1972~\cite{kugel2}, see also~\cite{kugel}, and is a direct generalization to
the case of orbital degeneracy of the usual
superexchange~\cite{anderson}.

A convenient mathematical way to describe orbital ordering is to
introduce operators $T_i$ of the pseudospin $\frac12$,
describing the orbital occupation, so that e.g.\ the state
$|T^z=\frac12\rangle$ corresponds to the occupied orbital
$|z^2\rangle$, and $|T^z=-\frac12\rangle$---to $|x^2-y^2\rangle$.
The first one corresponds to a local elongation of the
O$_6$-octahedra (distortion coordinate $Q_3>0$) and the second---to local
contraction $Q_3<0$~\cite{kanamori}.  The second degenerate
$E_g$-phonon which can also lift electronic $e_g$-degeneracy, $Q_2$,
corresponds to a pseudospin operator $T^x$.  One can describe an
arbitrary distortion and corresponding wave function by  linear
superpositions of the states $|T^z=+\frac12\rangle$ and
$|T^z=-\frac12\rangle$
\begin{equation}
\textstyle|\theta\rangle=\cos\frac\theta2|\frac12\rangle+\sin\frac\theta2|-\frac12\rangle
\label{eq1}
\end{equation}
where $\theta$ is an angle in $(T^z,T^x)$-plane.

The first, Jahn-Teller mechanism of the orbital ordering starts from
the electron--phonon interaction which in our case can be written in
the form
\begin{equation}
H=\sum g_{iq}[T^z_i(b^\dagger_{3q}+b^{\vphantom{\dagger}}_{3,-q})+
T_i^x(b^\dagger_{2q}+b^{\vphantom{\dagger}}_{2,-q})]+
\sum\omega_{\alpha q}b^\dagger_{\alpha q}b^{\vphantom{\dagger}}_{\alpha q}
\end{equation}
where $\alpha=2;3$ and $b^\dagger_3$ and $b^\dagger_2$ are the phonon
operators corresponding to $Q_3$ and $Q_2$ local modes.  Excluding
the phonons by a standard procedure, one obtains the orbital
interaction having the form of a pseudospin--pseudospin interaction
\begin{equation}
H_{\it eff}=\sum_{ij}J_{ij}^{\mu\nu}T_i^\mu T_j^\nu
\label{eq3}
\end{equation}
where
\begin{equation}
J_{ij}\sim\sum_q\frac{g_q^2}{\omega_q}e^{iq(R_i-R_j)}
\end{equation}
and $\mu,\nu=x,y$.  Due to different dispersion of different
relevant phonon modes, and to anisotropic nature of
electron--phonon coupling, the interaction (\ref{eq3}) is in general
anisotropic.

Similarly, the exchange mechanism of orbital ordering may be
described by the Hamiltonian containing the pseudospins $T_i$, but
also ordinary spins $\vec S_i$.  It can be derived starting from the
degenerate Hubbard model~\cite{kugel2}, and has schematically the
form
\begin{equation}
H=\sum_{ij}\bigl\{J_1\vec S_i\vec S_j+J_2(T_iT_j)+J_3(\vec S_i\vec S_j)(T_iT_j)\bigr\}.
\label{eq5}
\end{equation}
Here the orbital part $(T_iT_j)$, similar to (\ref{eq3}), is in
general anisotropic, whereas the spin exchange is Heisenberg-like.
In contrast to the Jahn-Teller induced interaction, the exchange
mechanism describes not only the orbital and spin orderings
separately, but also the coupling between them (last term
in~(\ref{eq5})).  This mechanism is rather successful in explaining
the spin and orbital structure in a number of
materials~\cite{kugel2,kugel}, including LaMnO$_3$ (for the latter
one has to invoke also the anharmonicity effects~\cite{kugel2}---see
also~\cite{khomskii2p}).

As to the electron--lattice interaction, typically one includes
mostly the coupling with the
local---i.e.\ optical---vibrations~\cite{halperin}.  However no less
important may be the interaction with the long-wavelength acoustical
phonons, or, simply speaking, with the elastic deformations.
Generally, when one puts an impurity in a crystal, e.g.\
replacing the small Mn$^{4+}$ ion in CaMnO$_3$ by the somewhat larger
Mn$^{3+}$ ion which in addition causes local lattice distortion due
to Jahn-Teller effect (i.e.\ we replace a ``spherical'' Mn$^{4+}$ ion
by an ``ellipsoidal'' Mn$^{3+}$), this creates a strain field which is
in general anisotropic and decays rather slowly, as
$1/R^3$~\cite{eshelby,khachaturyan}.  A second ``impurity'' of this
kind ``feels'' this strain, which leads to an effective long-range
interaction between them.  This can naturally lead to the spontaneous
formation of different superstructures in doped
materials~\cite{khomskii3}.  Thus, there may appear vertical or
diagonal stripes, even for non-Jahn-Teller systems.  In case of
manganites one can show that there appears an effective attraction
between e.g.\ $x^2$ and $y^2$-orbitals in $x$ and $y$-direction; this
immediately gives the orbital ordering of LaMnO$_3$-type shown in
fig.~3.  For $x=0.5$, assuming the checkerboard charge ordering, one
gets due to this mechanism the correct orbital ordering shown in
fig.~4~\cite{khomskii3}.  And for overdoped manganites one can
get either single or paired stripes, depending on the ratio of
corresponding constants:  One can show~\cite{halperin,khomskii3} that
for a diagonal pair like the ones in fig.~5, one gets an attraction
of the same orbitals $x^2$ and $x^2$ or $y^2$ and $y^2$, but
repulsion of $x^2$ and $y^2$. Thus, if one
takes into account only these nearest neighbour diagonal
interactions, the single stripe phase of fig.~5$a$ would be more
favourable than the paired stripes of the fig.~5$b$. However the
latter may in principle be stabilizes by more distant interactions
like those for a pair of Mn$^{3+}$ ions along $x$ and $y$-directions
in fig.~$5b$. Which state is finally more favourable, is now under
investigation.

\section{Ferromagnetic insulating phase at low doping}

As already mentioned in sec.~2, typically there exist a ferromagnetic
insulating region at low doping
($0.1<x<0.18$
for the LaSr system, $x<0.25$ for LaCa, $0.15<x<0.3$ for PrCa). The
problem is to explain the origin of the FI state in this case.
Apparently it should be connected with an orbital ordering of some
kind; but what is the specific type of this ordering, is largely
unknown.

The most complete data exist for the LaSr-system close to $x=1/8$. There
exists a superstructure in this system~\cite{yamada,klingerer}, and an orbital
ordering was detected in the FI phase in~\cite{endo}. Certain
orbital superstructure was also seen by the anomalous X-ray
scattering in  Pr$_{0.75}$Ca$_{0.25}$MnO$_3$~\cite{zimmermann2}. Both
these systems however were looked at at only one $k$-point [300],
which is not sufficient to uniquely determine the type of orbital
ordering.

Theoretically two possibilities were discussed in the
literature~\cite{mizokawa,mizokawa2}. First of all one can
argue that when one puts a Mn$^{4+}$ ion into Mn$^{3+}$ matrix, the
orbitals of all the ions surrounding the localized hole (Mn$^{4+}$)
would be directed towards it, see fig.~6$a$~\cite{mizokawa2,khaliullin}.
One can call such  state an orbital polaron.
%(32;  we with orbital polarons).
These polarons, which according to
Goodenough--Kanamori--Anderson rules would be ferromagnetic, can then
order e.g.\ as shown in fig.~6$b$ for $x=0.25$. The calculations
carried out in \cite{mizokawa2} show that this state is indeed
stable, and it corresponds to a ferromagnetic insulator. Thus it is a
possible candidate for a FI state at $x\simeq1/4$ e.g.\ in Pr--Ca
system.

However there exist an alternative possibility. The calculations
show~\cite{mizokawa} that similar state with ordered polarons is also
locally stable for  $x\simeq1/8$. But it turned out that the lower
energy is reached in this case by different type of charge and
orbital ordering, fig.~6$c$~\cite{mizokawa}: the holes are localised
only in every second $xy$-plane, so that one such plane containing
only Mn$^{3+}$ ions develops the orbital ordering of the type of
LaMnO$_3$, fig.~3, and the holes in the next plane concentrate in
``stripes'', e.g.\ along $x$-direction. This state also turns out to be
ferromagnetic, and the superstructure obtained agrees with the
experimental results of \cite{yamada} and \cite{klingerer} for
La$_{1-x}$Sr$_x$MnO$_3$, $x\simeq1/8$. One can think that the
situation can be also similar for $x\simeq1/4$ which would agree
with the data of~\cite{zimmermann2}. This type of the charge ordering
(segregation of holes in every second plane) may be favourable due
to an extra stability of the LaMnO$_3$-type orbital ordering,
strongly favoured by the elastic interactions, as discussed in sec.~3.

\section{Overdoped manganites}
Now I will qualitatively discuss the role of orbital degrees of
freedom in the overdoped regime, $x>0.5$. The main question is why in
this case the conventional double exchange, apparently responsible
for the formation of the ferromagnetic metallic state for $x\sim0.3$
-- $0.4$, does not lead to such a state in this case.

One reason may be the following. Usually we ascribe ferromagnetism in
doped systems to a tendency to gain kinetic energy by maximal
delocalization of doped charge carriers. These carriers are holes in
lightly doped manganites  $x\ll1$  , and electrons when we start
e.g.\ from CaMnO$_3$ and substitute part of Ca by La or other rare
earths, which corresponds to $x<1$ in La$_{1-x}$Ca$_x$MnO$_3$.

There exist an important difference between these two cases, however.
When we dope LaMnO$_3$, the orbital degeneracy in the ground state is
already lifted by orbital ordering, and in a first approximation
we can consider the motion of doped holes in a nondegenerate band.
Then all the standard treatment, e.g.\ of de~Gennes~\cite{degennes},
applies, and we get the FM state. However, when we start from the
cubic CaMnO$_3$, we put extra electrons into empty
{\it degenerate} $e_g$-levels, which form degenerate bands.
Therefore we have to generalize the conventional double-exchange
model to the case of degenerate bands. This was done
in~\cite{vandenbrink2}, and the outcome is the following: At
relatively low electron concentration ($x\simeq1$) the anisotropic
magnetic structures---C-type (chain-like) or A-type (ferromagnetic
planes stacked antiferromagnetically)---are stabilised, and only
close to $x\sim0.5$ do we reach the ferromagnetic state. The C-phase
occupies larger part of the phase  space. The resulting theoretical
phase diagram~\cite{vandenbrink2} is in surprisingly good agreement
with the properties of Nd$_{1-x}$Sr$_x$ MnO$_3$~\cite{kuwahara} in
which there exist the A-type ``bad metal'' phase for $0.52<x<0.65$
and C-phase for $x>0.65$.

Simple qualitative explanation of this tendency is the
following. When we start from CaMnO$_3$ with Mn${^4}$
($t_{2g}$)-ions and dope it by electrons, we put electrons into
$e_g$-bands. The maximum energy we can gain is to put these electrons
at the bottom of corresponding bands, so that one has to make these
bands as broad as possible. But due to a specific character of the
overlap of different $e_g$-orbitals in different directions, the
bottom of the bands coincides for different types of orbitals: one
can easily check that if we make all the orbitals e.g.\ $z^2$, the
energy $\epsilon(k)$ at the $\Gamma$-point $k=0$ will be the same as
for the bands made of ($x^2-y^2$)-orbitals. (Actually it is a
consequence of the degeneracy of $e_g$-orbitals in cubic crystals:
the symmetry at the $\Gamma$-point should coincide with the point
symmetry of local orbitals, i.e.\ at $k=0$ the energies of the
$z^2$-band and of the $(x^2-y^2)$ one, or of a band made of any
linear combinations thereof of the type~(1), should coincide.)

But according to the double-exchange model electrons can move only
if localised moments ($t_{2g}$-spins) of the corresponding
sites are ordered ferromagnetically (although without doping, in
CaMnO$_3$ ($x=1$), the magnetic ordering is antiferromagnetic
(G-type)). Now, if we make the band e.g.\ out of
($x^2-y^2$)-orbitals, the band dispersion would have the form
\begin{equation}
\epsilon(\vec k)=-2t(\cos k_x+\cos k_y)
\end{equation}
i.e.\ the electrons in this band move only in the $xy$-plane, but
there is no dispersion in the $z$-direction. Therefore to gain full
kinetic energy it is enough to make this plane ferromagnetic, and the
adjacent planes may well remain antiparallel to the first one. But
this is just the A-type magnetic structure (ferromagnetic planes
stacked antiferromagnetically).

Note that in this case the electron occupy predominantly
$(x^2-y^2)$-states (or $z^2$-states in case of C-type ordering).
Accordingly there will be corresponding lattice distortion
(compression along $c$-axis, $c/a<1$, for the A-type structure, and
$c/a>1$ for the C-type one). But I want to stress that these are
strictly speaking not the localized orbitals, but rather {\it bands}
of corresponding character.  Whether we should call it orbital
ordering, is a matter of convention (usually this terminology is
applied to the case of localized orbitals).  In any case, the feature
mentioned above (\cite{kanamori,khomskii2p}), that due to higher-order
effects, in particular lattice anharmonicity, only locally elongated
MeO$_6$-octahedra are observed in practice, is valid only for orbital
ordering of {\it localized} orbitals, and it is in general not true
for the band situation considered here.

Thus the double exchange via degenerate orbitals may quite naturally
lead to anisotropic magnetic structures (A-type or C-type): we gain
by that the full kinetic energy without being forced to sacrifice all
the exchange interaction of localised electrons (part of the bonds
remain antiferromagnetic). Which particular state will be stable at
which part of the phase diagram, is determined by the competition
between these terms, kinetic energy vs exchange energy, with the
electron energy depending on the band filling sensitive to the density of
states for the corresponding band.

There are several factors which can complicate this picture. Thus, one
may in principle get in this case canted states, and not the fully
saturated A- or C-type structures~\cite{vandenbrink2,soloviev2}. There may appear
also inhomogeneous phase-separated states. The possibility of the
charge ordering (e.g.\ in the form of stripes) was also not considered
in this treatment. But altogether it shows that the conventional
double-exchange picture should be modified if double-exchange goes
via degenerate orbitals, and overall tendency which results due to
this is that not the simple ferromagnetic state, but more complicated
magnetic structures may be stabilised, which agrees with the general
tend observed in the experiment. This factor may be important in
explaining strong qualitative asymmetry of the phase diagram of
manganites for  $x < 0.5$ (underdoped) and $x > 0.5$ (overdoped)
regimes.

\section{Orbital ordering in ferromagnetic metallic phase?}
Finally we go over to the most important phase---that of optimal
doping, $x\simeq0.3$ -- $0.5$. In most cases the systems in this
doping range at low temperatures are ferromagnetic and metallic,
although the residual resistivity is usually relatively large.

Now, the question is: what are the orbitals doing in this phase~?
Experimentally  one observes that the macroscopic Jahn-Teller ordering
and corresponding lattice distortion is gone in this regime. La--Ca
system remains orthorhombic in this concentration range, but it is due
to the tilting of the O$_6$ octahedra, octahedra themselves being
regular (all the Mn--O distances are the same). The structure of the
La--Sr manganites in this regime is rhombohedral, but again all the
Mn--O distances are equal. Moreover even the local probes such as
EXAFS or PDF (pair distribution function analysis of
neutron scattering)~\cite{louca}, which detect local
distortions above and close to $T_c$, show that for $T\to0$ they
completely disappear, and MnO$_6$-octahedra are regular even locally.

What happens then with the orbital degrees of freedom~? There are
several possibilities. One is that in this phase the system may
already be an ordinary metal, electronic structure of which is
reasonably well described by the conventional band theory. In this
case we should not worry about orbitals at all: we may have band
structure consisting of several bands, some of which, not necessarily
one, may cross  Fermi-level, and we should not speak of orbital
ordering in this case, the same as we do not use this terminology
and do not worry about orbital ordering in metals like Al or Nb which
often have several bands at the Fermi-level.

If however there exist strong electron correlations in our system
(i.e.\ the Hubbard's on-site repulsion $U$ is bigger that the
corresponding bandwidth)---one should worry about it. The orbital
degrees of freedom should then do
something. There exists then two options. One is that the ground
state would still be disordered due to quantum fluctuations, forming
an orbital liquid~\cite{ishihara}, similar in spirit to the RVB state of the spin
system (we can speak of the pseudospin RVB state). This is of course
in principle possible. I however see some problems with this picture.
One is that typically  the orbital (pseudospin) interaction is not
Heisenberg-like, but rather anisotropic, see Eqs.\ (\ref{eq3}),~(\ref{eq5}).
Whether it is good or bad for the orbital liquid state,
remains at present highly
controversial~\cite{oudovenko,feiner}. Another factor is that
due to a rather strong Jahn-Teller interaction (or pseudospin--phonon
coupling) one may expect strong suppression of quantum fluctuations
by the polaron effects. And finally, an argument against this picture
is the already mentioned experimental observation that the lattice
structure is undistorted even on a local level (these two latter
factors may however be explained if we assume that the orbital
quantum fluctuations are very fast, i.e.\ occur at a time scale  much
shorter that the phonon times $\sim\hbar/\omega_D$).

There exists however yet another, alternative possibility: there
may in principle occur in this case an orbital ordering of a novel
type, without any lattice distortion, involving not the orbitals of
the type~(\ref{eq1}), but the {\it complex} orbitals---linear
superpositions of the basic orbitals $z^2$ and $(x^2 - y^2)$ with the
complex coefficients, e.g.
\begin{equation}
|\pm\rangle=\frac{1}{\sqrt2}\Bigl(|z^2\rangle\pm
i|x^2-y^2\rangle\Bigr).
\label{eq7}
\end{equation}
This possibility was first suggested in~\cite{khomskii4,vandenbrink3}
and explored in~\cite{maezono}; independently similar conclusion
was reached a bit later in~\cite{takahashi}.

Why can such state be favourable  and what are its properties~? My
initial arguments were based on the analogy with the double-exchange
model and with the well-known phenomenon of Nagaoka's
ferromagnetism~\cite{nagaoka}. When we introduce holes into strongly
correlated Hubbard system, $U\gg t$, $n<1$, the tendency to
gain kinetic energy forces the system to become ferromagnetic, at
the expense of losing the antiferromagnetic exchange energy of
localised spins.

One can show that pseudospins play the role similar to
ordinary spins: one gains kinetic energy when one makes the orbitals
ordered. For very low doping $x\ll1$, when all the holes are at the
bottom of the band, one would get a ferro.\ ordering of orbitals. For
finite $x$, due to a finite band filling and different behaviour of
the density of states for different bands, some other
types of orbital ordering may become preferable, e.g.\ the staggered
(antiferro.)\ orbital ordering---this would be determined by which
structure minimizes total band energy for a given band filling 
(this is the same factor which determines the stability of one
or another magnetic phases in the situation considered in
sec.~5~\cite{vandenbrink2}).

Why then the complex orbitals (\ref{eq7}) and not the conventional
real combinations (\ref{eq1})~? One can show that, similar to the
treatment of sec.~5, the bottom of the band for the complex orbitals
(\ref{eq7}) will be {\it exactly the same} as for any real combination
(\ref{eq1}), so that in this sense the orbitals (\ref{eq7}) are at least not worse
than the conventional ones. On the other hand, they may be better
from the point of view of the exchange interaction.

One can show that for undoped systems the ordinary orbitals are
always better: they are stabilised by both the
exchange and Jahn-Teller interactions (\ref{eq5}), (\ref{eq3}),
and formally it is reflected in the fact that these effective
Hamiltonians contain only pseudospin operators $T^z$,
$T^x$. One can easily show however that the state (\ref{eq7}) is an
eigenstate of the third operator
$T^y=\frac12\Bigl({0\atop-i}\;{i\atop0}\Bigr)=\frac12\sigma_y$, where
$\sigma_y$ is the corresponding Pauli matrix. As these operators do
not enter the Hamiltonians (\ref{eq3}), (\ref{eq5}), the
corresponding pseudospin (orbital) ordering for undoped systems would
be the one in ($T^z,T^x$)-sector, i.e.\ it would be the ordering of
real orbitals.

But just the fact that the $T^y$-operators do not enter the
pseudospin exchange interactions (\ref{eq3}), (\ref{eq5}), which are typically
``antiferromagnetic'' (antiferroorbital!) tells us that if we
{\it force} our system to change the type of orbital ordering,
e.g.\ making it ``ferromagnetic'', so as to gain maximum kinetic
energy---then it may be favourable to make it $T^y$-ferro.orbital
ordering: we gain by that {\it the same} kinetic (band) energy,
and lose {\it less exchange} energy.

The real Hartree-Fock calculations carried out
in~\cite{vandenbrink3,maezono} indeed confirm that there
exist conditions at which the ordering of complex orbitals is
energetically preferable to that of the real ones. Thus, for the
realistic values of parameters (electron hopping $t$, Hubbard
interaction~$U$) the staggered ordering of complex orbitals may be
realized at doping level $x\sim0.35$ -- $0.4$ --- just in the most
important region of the phase diagram of manganites (the
ferro.ordering of complex orbital could be realized at larger values
of $U/t$ at small doping~\cite{khomskii4,vandenbrink3}).

The properties of this novel type of orbital ordering were
investigated in~\cite{khomskii4,vandenbrink3}. One may easily see that the 
distribution of the electron density in this state is the same in all three
directions, $x$, $y$ and~$z$. Thus this ordering does not induce any
lattice distortion---the MnO$_6$ octahedra remain regular, and the
system is cubic (if we ignore tilting of the octahedra). On the other
hand, the state (\ref{eq7}), as always is the case with complex wave
functions, breaks time-reversal invariance, i.e.\ this state is in
some sense magnetic. One can show however that the magnetic dipole
moment in this case is zero---it is well known that the orbital
moment is quenched in $e_g$-states (these states are actually
$|l^z=0\rangle$ and $\frac{1}{\sqrt2}(|2\rangle + \mathopen{|}-2\rangle)$
states of the $l=2$ $d$-orbitals). Similarly,
magnetic quadrupole moment is also zero, by parity arguments. The
first nonzero moment in this state is a magnetic octupole. Indeed,
the actual order parameter in this case is the average

\begin{equation}
\eta=\langle M_{xyz}\rangle=\langle{\rm S}L_xL_yL_z\rangle\neq0
\label{eq8}
\end{equation}
where $l_\alpha$ are the components of the orbital moment $l=2$ of
$d$-electrons, and S means the symmetrization. This operator is
actually proportional to the $T^y$-operator of pseudospin, i.e.\ the
order parameter of this type of orbital ordering is indeed
$\eta=\langle T^y\rangle$.

One can visualize this state as the one in which there exist orbital
currents at each unit cell. But these currents have rather high
symmetry, so that the resulting magnetic fields are of octupole
character, see fig.~7.

One can analyze some other properties of the state (\ref{eq7}) with
octupole ordering~\cite{khomskii4,vandenbrink3}). The main problem is to 
find an experimental probe which could directly check the existence of this
octupole ordering. This is indeed not easy, but if successful, such
experiments would allow us to verify whether such novel state is
indeed realized in optimally doped ferromagnetic metallic manganites.
If correct, this would mean that this state is perfectly ordered not
only with respect of spins but also  as to the orbitals, but with
``strange'' orbitals~(\ref{eq7}). (Relatively large residual
resistivity in this case may be due to the (small scale and possibly
dynamic) phase separation.) If not, this would probably mean that
the ground state in this case is a quantum orbital liquid. In both
cases we lose (or rather do not gain) Jahn-Teller energy. Whether
quantum effects are sufficiently strong to overcome the factors
discussed above which stabilize complex orbitals, is still not clear
at present.

\section{Orbital polarons and properties of manganites at finite
temperatures.}
In this last section I want to shortly discuss some of
the issues which are rather actively investigated nowadays---those of
the short-range orbital correlations and local Jahn-Teller effect at
finite temperatures, in situations where there is no long-range
orbital ordering. Specifically, these effect are often observed
above and close to the phase transitions. As we already saw, there
exists an orbital ordering of one or another type in manganites in
most of the cases. Consequently, one should expect that there will
exist at least local orbital correlations in disordered phases above
corresponding $T_c$'s. Such correlations were indeed observed e.g.\
in~\cite{zimmermann}.

The most important however are the recent observations  that such
correlations exist above $T_c$ and are enhanced in approaching $T_c$
(but rapidly disappear below it) even in optimally doped manganites
with $x\sim 0.3$ -- $0.4$ \cite{louca,vasiliudoloc}---in situation 
where there exists no orbital ordering at low
temperatures. Actually the idea of the possible importance of such
correlations in the paramagnetic phase of the CMR manganites was
first put forth by Millis and his coworkers already in 1995~\cite{millis};
they argued that the double exchange alone is not sufficient to
explain transport properties of manganites in this regime, and
suggested that they may be largely dominated by the Jahn-Teller
interaction.

The real direct indications that it may indeed be the case were
obtained only recently; the most spectacular one is that the
intensity of the diffuse neutron scattering attributed to Jahn-Teller
polarons closely follows the temperature dependence of the
resistivity~\cite{vasiliudoloc} (it grows with  decreasing temperature,
has a maximum at $T_c$ and rapidly disappears for $t<T_c$).

There are now many other experiments which are interpreted in terms
of these Jahn-Teller polarons (although this very notion is often not
well defined). This is quite a big field in itself, and I have no
space to discuss it here in details. Suffice it to say that the
orbital degrees of freedom apparently play very important role in
many properties of manganites not only in phases in which there
exists an orbital ordering, but also in disordered states.

\section{Conclusions}
In conclusion I can only repeat that orbital effects play very
important role in the physics of manganites, and also in many other
transition metal oxides~\cite{kugel,nagaosa}. Together with
charge and spin degrees of freedom they determine all the  rich
variety of the properties of manganites in different doping regions.
Orbital effects also play very important role in disordered phases,
determining to a large extent their transport and other properties.

An important recent achievement in this field is the development of
the method to directly study orbital ordering using the anomalous
resonant X-ray scattering, initiated by the pioneering work of
Murakami et al.~\cite{murakami}. This method was successfully
applied to a number of problems in
manganites as well as
to several other systems. And although there is still a controversy
as to the detailed microscopic explanation of these
observations~\cite{ishihara2,elfimov}, this method will be
definitely of great use in the future.

The last point I want to mention is that until now I discussed in
this paper mostly the static (ground state) properties connected with
orbital ordering. However each time we have certain ordering in
solids, there should appear corresponding excitations in them. In
our case these excitations---we may call them orbitons---were first
discussed shortly in~\cite{kugel2,kugel} (and in more details in the PhD
thesis by K.~I.~Kugel in 1975) and recently were studied
theoretically in several papers, e.g.\
in~\cite{ishihara3,vandenbrink4}. One of the problems which,
in my opinion, could have made an experimental observation of these
excitations difficult, is the usually rather strong Jahn-Teller
coupling of orbital degrees of freedom with the lattice distortions.
I was afraid that it could make very difficult, if not impossible, to
``decouple'' orbitons from phonons. And indeed the experimental
efforts to observe orbitons were unsuccessful for many years. The
breakthrough was made only recently when the group of Y.~Tokura
managed to observe orbital excitations by Raman scattering in
untwinned single crystals of LaMnO$_3$~\cite{saitoh}. And although
many questions here still remain unclear, this work will definitely
open a new chapter in the study of orbital effects in oxides, in
particular in manganites. Thus the field of orbital physics is still
capable of producing important new results, and sometimes---surprises.

In conclusion I want to thank many of my colleagues with whom I had a
pleasure to collaborate and to discuss the exciting questions of
orbital physics. Among many good friends and colleagues I would like
to single out three: K.~I.~Kugel with whom we started long
collaboration in this field already quite a while ago and continue it
until now, and my recent collaborators J.~van~den~Brink and
G.~A.~Sawatzky who contributed a lot to the recent development of this
field. And, once again, I want to use this opportunity to congratulate
E.~M\"uller-Hartmann with his jubilee and to wish him many fruitful and
happy years.

\pagebreak

\section*{Figure Captions}

\begin{description}
\item{Fig.~1. \ }Phase diagram of La$_{1-x}$Ca$_x$MnO$_3$
(after to S.-W.~Cheong).  O---orthorhombic phase with rotated regular octahedra; 
O$'$---orthorhombic phase with Jahn-Teller distortions,

\item{Fig.~2. \ }The splitting of $3d$-levels in a cubic crystal field
(regular MnO$_6$-octahedron).  Electron occupation of $4d$-electrons
in Mn$^{3+}$ is shown by arrows

\item{Fig.~3. \ }The orbital ordering of LaMnO$_3$.  Arrows show shifts of oxygen ions

\item{Fig.~4. \ }Charge, orbital and spin ordering in the basal
($xy$)-plane of manganates at $x=0.5$.  Arrows denote the spin ordering. 
The spin zigzags are shown by thick lines

\item{Fig.~5. \ }($a$)~Single stripes (``Wigner crystal'') and ($b$)
paired stripes, or bistripes in La$_{1-x}$Ca$_x$MnO$_3$ for $x=\frac23$.  
O---ions Mn$^{4+}$; $8$, $\infty$---ions Mn$^{3+}$ with the corresponding orbitals

\item{Fig.~6. \ }Orbital polarons and possible types of orbital
ordering in low doped manganites:
($a$)~Orbital polaron close to a Mn$^{4+}$ ion; ($b$)~Ordering of
orbital polarons for $x=0.25$ in a bcc-lattice; ($c$)~An alternative
charge and orbital ordering, obtained for $x=1/8$ in~\cite{mizokawa}.  Notations
are the same as in figs.~4,~5.  Shaded lines---``stripes'' containing holes.

\item{Fig.~7. \ }The distribution of magnetic field around each Mn in
the ordered state with the complex orbitals~(\ref{eq7}). The $+$ and
$-$ signs show the direction of the magnetic field (outward and
inward).  One sees that the local symmetry axes are 4 cube diagonals
[111], in accordance with~(\ref{eq8})

\end{description}

\pagebreak
\Large
%\null\vfil\centerline{\epsfbox{fig1.eps}}\bigskip\bigskip\centerline{Fig.~1}\vfil\break

%\null\vfil\centerline{\epsfbox{fig2.eps}}\bigskip\bigskip\centerline{Fig.~2}\vfil\break

%\null\vfil\centerline{\epsfbox{fig3.eps}}\bigskip\bigskip\centerline{Fig.~3}\vfil\break

%\null\vfil\centerline{\epsfbox{fig4.eps}}\bigskip\bigskip\centerline{Fig.~4}\vfil\break

%\null\vfil\centerline{\epsfbox{fig5.eps}}\bigskip\bigskip\centerline{Fig.~5}\vfil\break

%\null\vfil\centerline{\epsfbox{fig6.eps}}\bigskip\bigskip\centerline{Fig.~6}\vfil\break

%\null\vfil\centerline{\epsfbox{fig7.eps}}\bigskip\bigskip\centerline{Fig.~7}\vfil\break

\begin{figure}
\epsfxsize=10cm
\centerline{\epsffile{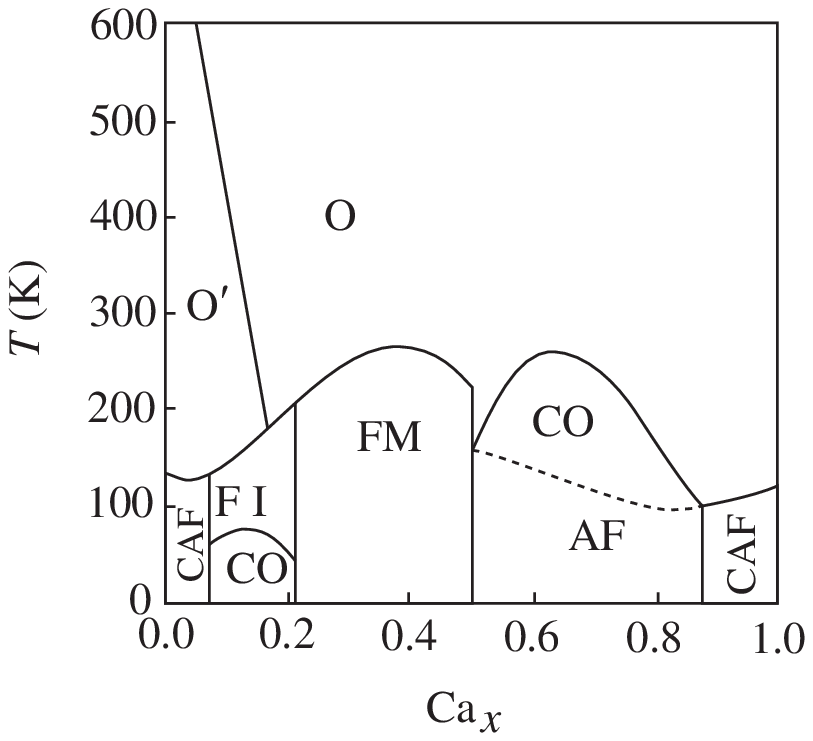}}
\centerline{Fig.~1}
\end{figure}

\begin{figure}
\epsfxsize=10cm
\centerline{\epsffile{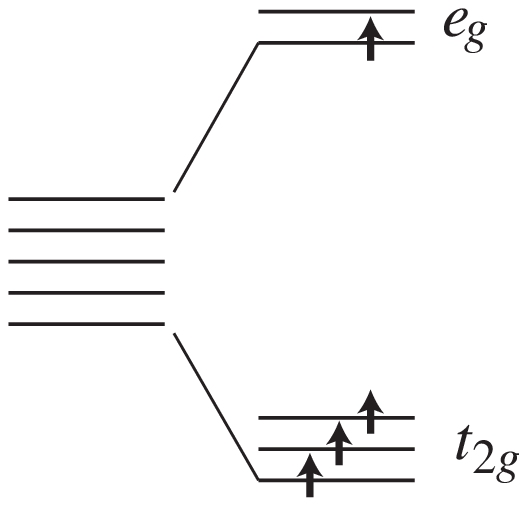}}
\centerline{Fig.~2}
\end{figure}

\begin{figure}
\epsfxsize=10cm
\centerline{\epsffile{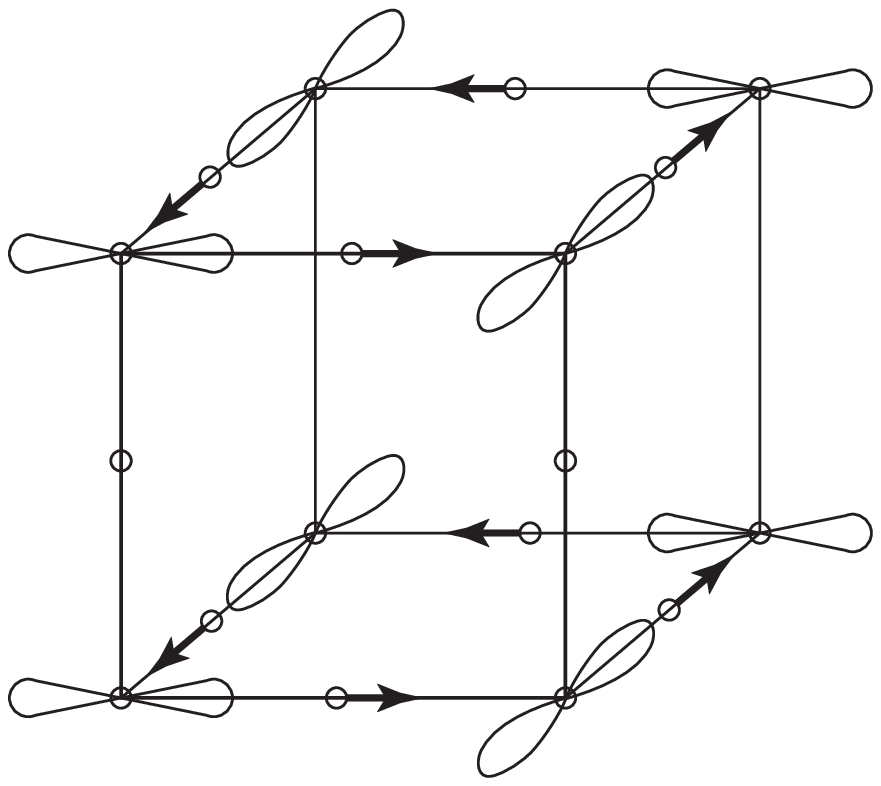}}
\centerline{Fig.~3}
\end{figure}

\begin{figure}
\epsfxsize=10cm
\centerline{\epsffile{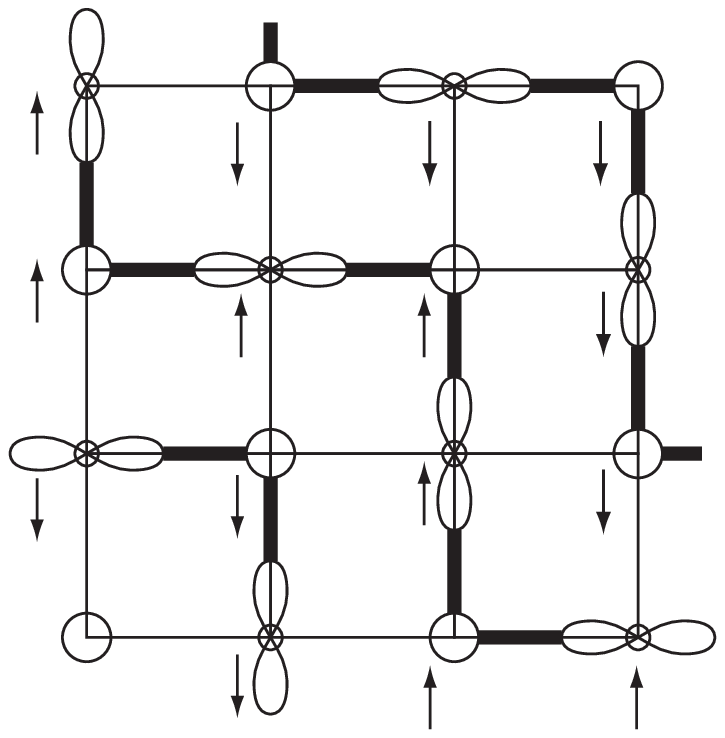}}
\centerline{Fig.~4}
\end{figure}

\begin{figure}
\epsfxsize=8cm
\centerline{\epsffile{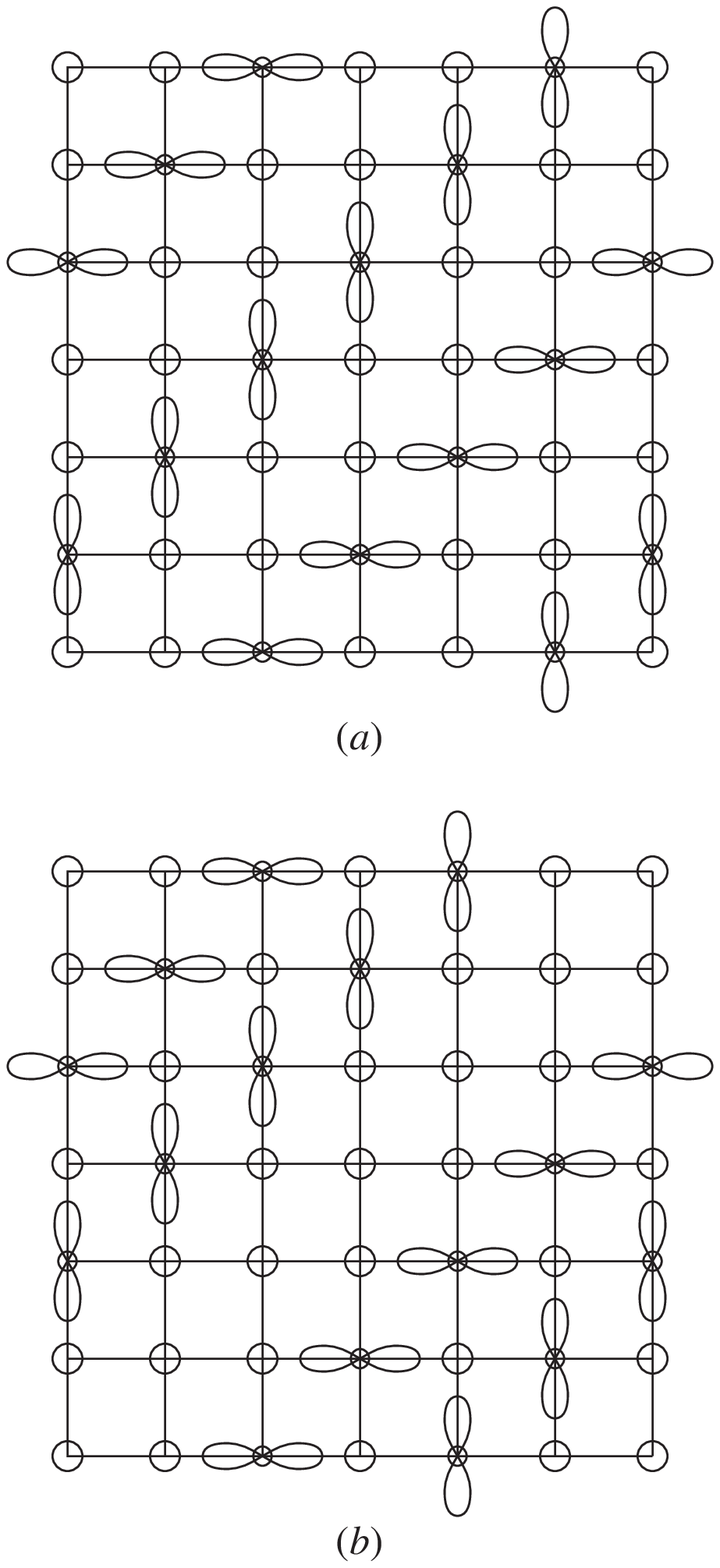}}
\centerline{Fig.~5}
\end{figure}

\begin{figure}
\epsfxsize=15cm
\centerline{\epsffile{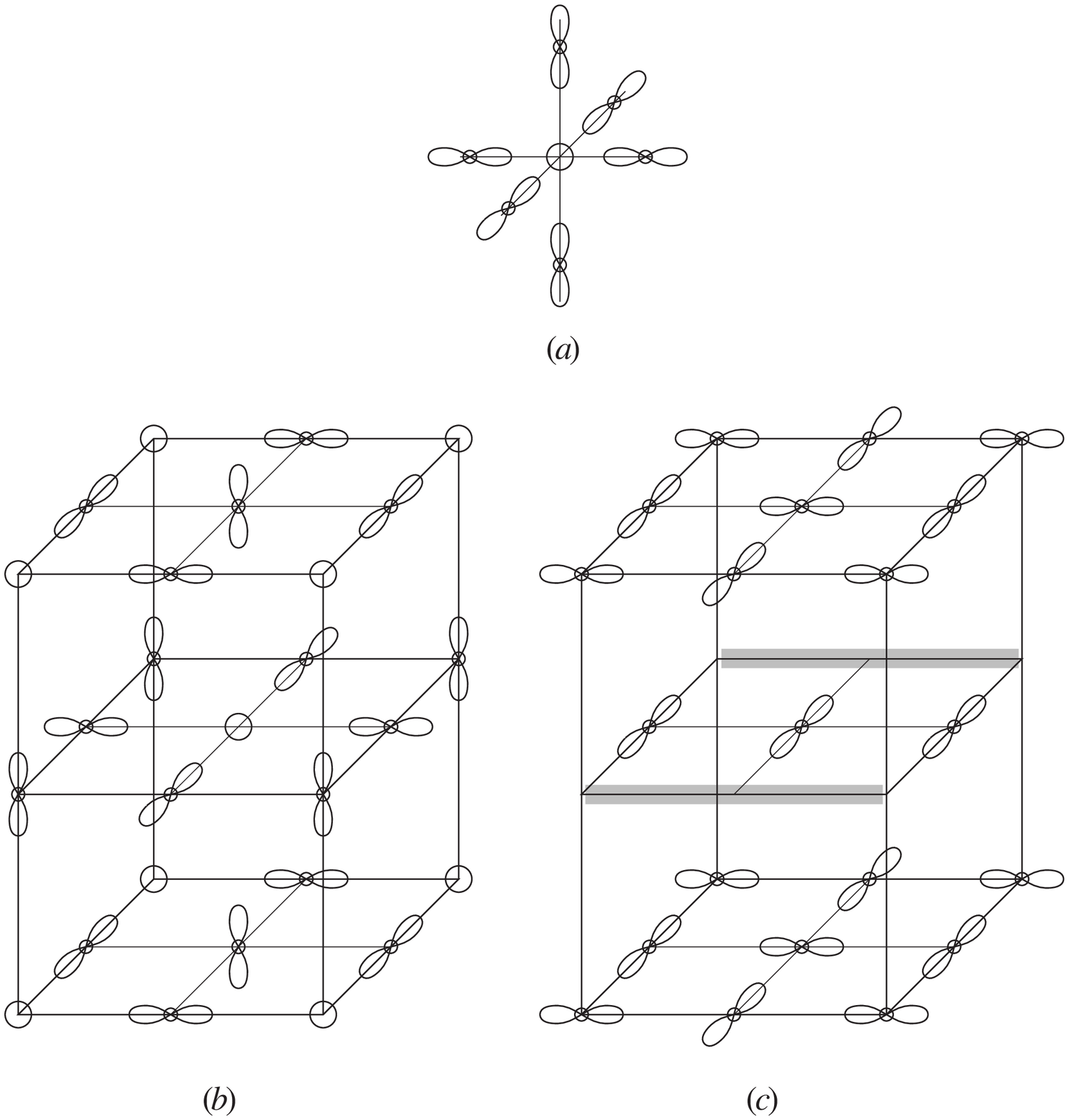}}
\centerline{Fig.~6}
\end{figure}

\begin{figure}
\epsfxsize=10cm
\centerline{\epsffile{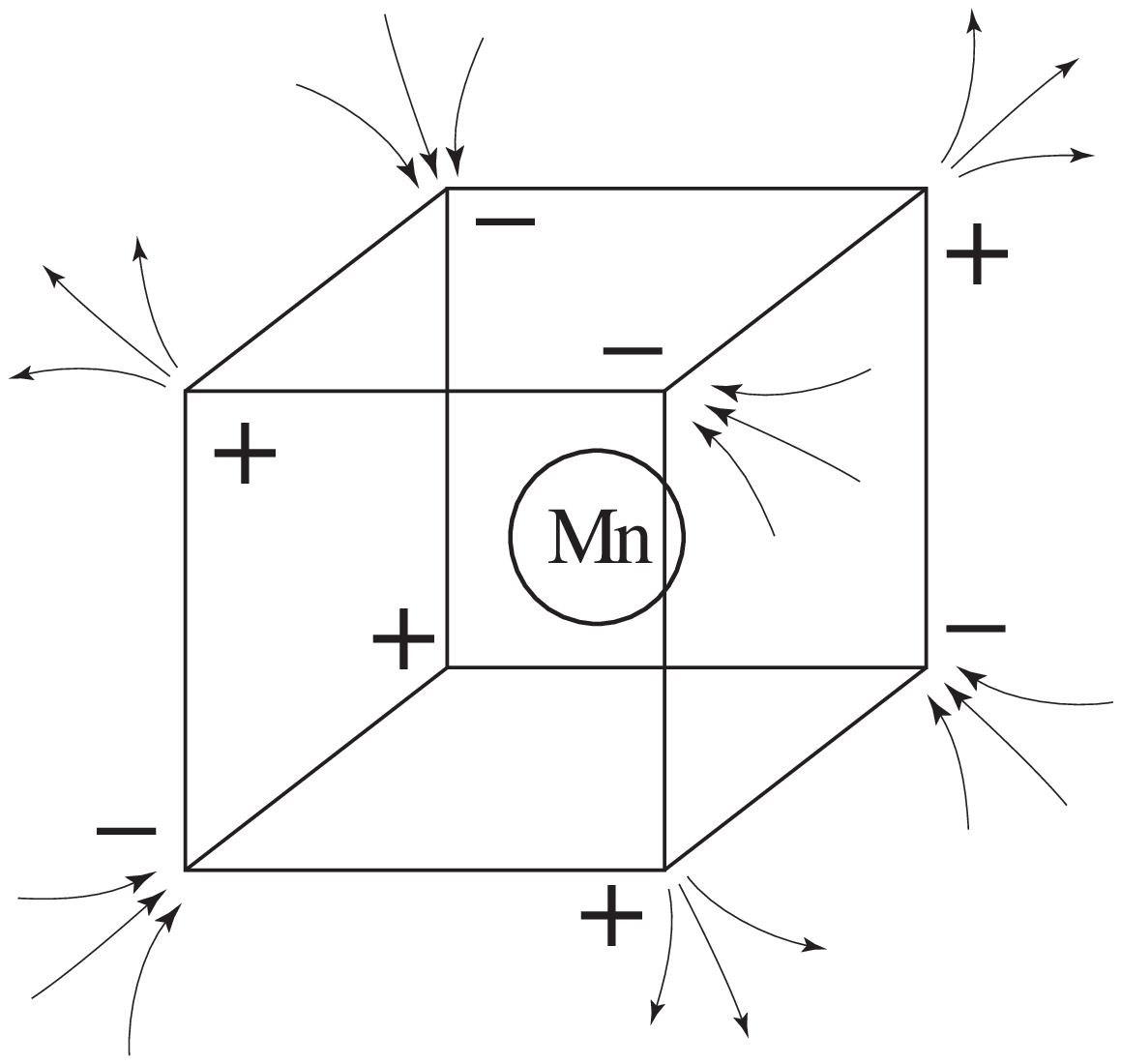}}
\centerline{Fig.~7}
\end{figure}

\end{document}